\documentclass[a4paper, oneside, twocolumn, notitlepage, 10pt]{extarticle_ecoc}
\usepackage{ecoc}

\begin{document}
\selectlanguage{english}    % Standard Language

%-------------------------------------------------- Title -----------------------------------------------------%

\title{423.7 + 426.5 Tb/s GMI Bi-Directional HCF Transmission}%

%------------------------------------------------- Authors-----------------------------------------------------%

\author{
    J.~Yang\textsuperscript{(1)}, R.~Aparecido\textsuperscript{(1)}, E.~Sillekens\textsuperscript{(1)}, R.~Sohanpal\textsuperscript{(1)}, M.~Jarmolovi\v{c}ius\textsuperscript{(1)}, Z.~Gan\textsuperscript{(1)}, 
    Y.~Hong\textsuperscript{(2)}, \\M.~Kamalian-Kopae\textsuperscript{(2)}, A.~Ali\textsuperscript{(2)}, S.~Bakhtiari~Gorajoobi\textsuperscript{(2)}, R.\,S.~Lu\'{i}s\textsuperscript{(3)}, D.~Orsuti\textsuperscript{(3)}, 
    A.~Donodin\textsuperscript{(4)}, \\
    V.~Mikhailov\textsuperscript{(5)}, J.~Luo\textsuperscript{(5)}, D.\,J.~DiGiovanni\textsuperscript{(5)}, 
    N.~Fontaine\textsuperscript{(6)}, L.~Dallachiesa\textsuperscript{(6)}, M.~Mazur\textsuperscript{(6)}, R.~Ryf\textsuperscript{(6)}, \\H.~Chen\textsuperscript{(6)}, D.~Neilson\textsuperscript{(6)},
    I.\,D.~Phillips\textsuperscript{(4)}, W.~Forysiak\textsuperscript{(7)}, S.\,K.~Turitsyn\textsuperscript{(4)}, H.~Furukawa\textsuperscript{(3)}, \\J.~Gaudette\textsuperscript{(2)}, D.\,J.~Richardson\textsuperscript{(2)}, B.\,J.~Puttnam\textsuperscript{(2)}, R.\,I.~Killey\textsuperscript{(1)}, and P.~Bayvel\textsuperscript{(1)}
    \vspace{-1em}
}

\maketitle                  % Create title and author

%------------------------------------------ Description of Authors ----------------------------------------------%

\begin{strip}
    \begin{author_descr}

        \textsuperscript{(1)} Optical Networks Group, UCL (University College London), London, UK,
        \textcolor{blue}{\uline{jiaqian.yang.18@ucl.ac.uk}}

        \textsuperscript{(2)} Microsoft Azure Fiber, SO51 9DL Romsey, UK

        \textsuperscript{(3)} National Institute of Information and Communications Technology (NICT), Koganei, Tokyo, Japan

        \textsuperscript{(4)} Aston Institute of Photonic Technologies, Aston University, B4 7ET Birmingham, UK

        \textsuperscript{(5)} Lightera Laboratories, NJ 08873 USA

        \textsuperscript{(6)} Nokia Bell Labs, NJ 07974, USA

        \textsuperscript{(7)} Smart Internet Lab, University of Bristol, BS8 1UB Bristol, UK
\vspace{-.5em}
    \end{author_descr}
\end{strip}

% \setstretch{1.1}
%-------------------------------------------------- Footnote -------------------------------------------------------%
\renewcommand\footnotemark{}
\renewcommand\footnoterule{}
%\let\thefootnote\relax\footnotetext{text}

%-------------------------------------------------- Abstract ---------------------------------------------------------%

\begin{strip}
    \begin{ecoc_abstract}
        % NOTE: Don't use a blank line here but start abstract right away to avoid an extra line break
        We demonstrate OESCL-band same-wavelength bi-directional transmission over 60\,km HCF with 42.5\,THz bandwidth, achieving GMIs comparable with the highest unidirectional SMF data-rates in both directions, with an aggregate of 423.7\,+\,426.5\,Tb/s.  ©2026 The Author(s) 
        \vspace{-.5em}
    \end{ecoc_abstract}
\end{strip}

%-------------------------------------------------- Introduction Section -------------------------------------------------------%

\section{Introduction}
Ultra-wideband (UWB) wavelength-division multiplexing (WDM) technologies have been widely explored to address the ever-growing optical network throughput demands. To date, transmission bandwidths exceeding 37\,THz and 42\,THz have been demonstrated both in laboratory experiments ~\cite{puttnam2024_402} and deployed fibre links~\cite{luis2026_450}, respectively, with throughputs exceeding 400\,Tb/s, nearly exhausting the available low-loss transmission window of conventional optical fibres. Several approaches have been proposed to further advance UWB systems capacity including hybrid single- and multi-mode systems~\cite{luis2025_430} and anti-resonant hollow-core fibre (HCF)~\cite{petrovich2025broadband}. The latter 
%throughputs exceeding 400\,Tb/s have been demonstrated over 50 km G.652 fibre \cite{puttnam2024_402} and 10\,km G.654 fibre with multi-mode O-band signals~\cite{luis2025_430}, with high data-rates  also demonstrated over multi-span~\cite{puttnam2024_339} and deployed fibre links~\cite{soma2024multiband,yang2025_300}. 
has emerged as an attractive transmission medium, with low latency, loss, and nonlinearity, including negligible stimulated Raman scattering (SRS), which must otherwise be carefully managed in UWB silica-fibre systems. These properties lead to improved signal-to-noise ratios (SNRs) and extended transmission reach~\cite{ali2025unrepeated,ge2025_1,hong2025real}. In addition, HCFs exhibit much lower Rayleigh backscattering (RB)~\cite{michaud-belleau2021backscattering}. This allows bi-directional (Bi-Di) transmission, potentially doubling the capacity of a fibre-optic cable without increasing fibre count. %cable density. %The absence of Rayleigh backscattering means that counter-propagating signals do not need to occupy different wavelength sets~\cite{erkilincc2017bidirectional} or optical bands~\cite{yang2026_135} to avoid in-band crosstalk from the back-scattered light, and neither require signal processing to cancel this crosstalk ~\cite{obara2007bidirectional,chen2024overcoming}. 
Bi-Di HCF transmission has been reported using C+L-band~\cite{ge2024first,ge2025field} and S+C+L-band~\cite{liu2024_502,zhang2025_273,liu2026tb,li2026beyond} systems with bandwidths of up to 20\,THz, limited by the 2\textsuperscript{nd} window HCF design. However, to date, the use of transmission bandwidths comparable to silica single-mode fibres (SMFs) ($>$35\,THz), which is only possible in HCFs using the fundamental transmission window~\cite{petrovich2025broadband}, has not been investigated.

\newcommand{\datapoint}[5]{
    \addplot[color=#4,only marks, mark=*, mark size=2pt,
    point meta=explicit symbolic, 
    nodes near coords={\cite{#1}#5},
    nodes near coords style={inner sep=1pt, anchor=#3},forget plot,
    ] coordinates {#2};
    }
    \newcommand{\datapointgminocite}[4]{
    \addplot[color=#4,only marks, mark=o, mark size=2pt, line width=1pt,
    point meta=explicit symbolic, 
    nodes near coords style={inner sep=1pt, anchor=#3},forget plot,
    ] coordinates {#2};
    }
    \newcommand{\datapointsqgmi}[5]{
    \addplot[color=#4,only marks, mark=square, mark size=2pt, line width=1pt,
    point meta=explicit symbolic, 
    nodes near coords={\cite{#1}},
    nodes near coords style={inner sep=1pt, anchor=#3},forget plot,
    ] coordinates {#2};
    }
    \newcommand{\datapointsq}[5]{
    \addplot[color=#4,only marks, mark=square*, mark size=2pt, line width=1pt,
    point meta=explicit symbolic, 
    nodes near coords={\cite{#1}},
    nodes near coords style={inner sep=1pt, anchor=#3},forget plot,
    ] coordinates {#2};
    }
    
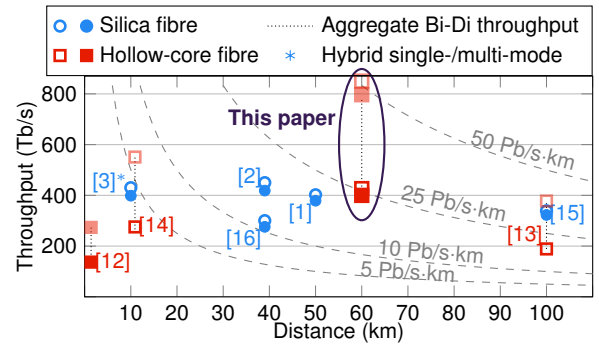
\begin{figure}[t]
    \centering
    % \vspace{-1em}
    \begin{tikzpicture}\footnotesize
      \hypersetup{hidelinks}
      \begin{groupplot}[
          	group style={
    		group size=1 by 1,
    		x descriptions at=all,
    		y descriptions at=all,
    		vertical sep=0em,
    	},
      % \begin{axis}[
    height=4.5cm,
    width=8.3cm,
    ymajorgrids=true,
    clip=false,
    ylabel style={at={(axis description cs:.07,.5)}},
    ]
    
    \nextgroupplot[
        xlabel={Distance (km)},
        ylabel={Throughput (Tb/s)},
        ymax=870,
        ymin=0,
        xmin=0,
        xmax=110,
        ytick={0,200,400,600,800},
        yticklabels={,200,400,600,800},
        xtick={0,10,...,110},
        xticklabels={,10,20,...,100,},
        legend style={column sep=1pt, fill=none, text opacity = 1, anchor=south east,at={(axis cs: 110,870)}},
        legend cell align={left},
        legend columns=3,
        xlabel style={at={(axis description cs:.5,.1)}},
        clip=false,
    ]

    \node[gray,rotate=-14] at (95,580) {\footnotesize 50 Pb/s$\cdot$km};
    \node[gray,rotate=-12] at (80,370) {\footnotesize 25 Pb/s$\cdot$km};
    \node[gray,rotate=-5] at (75,180) {\footnotesize 10 Pb/s$\cdot$km};
    \node[gray,rotate=-4] at (70,100) {\footnotesize 5 Pb/s$\cdot$km};

    \addplot[dashed, gray, domain=60:110, samples=100, forget plot]
    {50000/x};		
    \addplot[dashed, gray, domain=30:110, samples=100, forget plot]
    {25000/x};		
    \addplot[dashed, gray, domain=12:110, samples=100, forget plot]
    {10000/x};
    \addplot[dashed, gray, domain=6:110, samples=100, forget plot]
    {5000/x};

    \addplot[uclred, only marks,mark=square*, mark options={mark size = 2.4,line width=1pt},forget plot,
    point meta=explicit symbolic, 
    nodes near coords style={
        inner sep=2pt,
        anchor=north west,
    },
    ] coordinates {(60,399.85)};
    \addplot[uclred, only marks,mark=square, mark options={mark size = 2.4,line width=1pt},forget plot,
    point meta=explicit symbolic, 
    nodes near coords style={
        inner sep=2pt,
        anchor=north west,
    },
    ] coordinates {(60,426.51)};
    \addplot[uclred, only marks,mark=square*, mark options={mark size = 2.4,line width=1pt},opacity=0.5,forget plot,
    point meta=explicit symbolic, 
    nodes near coords style={
        inner sep=2pt,
        anchor=north west,
    },
    ] coordinates {(60,796.77)};
    \addplot[uclred, only marks,mark=square, mark options={mark size = 2.4,line width=1pt},opacity=0.5,forget plot,
    point meta=explicit symbolic, 
    nodes near coords style={
        inner sep=2pt,
        anchor=north west,
    },
    ] coordinates {(60,850.24)};
    
    \draw[densely dotted](axis cs:60,420)--(axis cs:60,780);
    \draw[color=ucldpurple,thick] (axis cs:60,600) ellipse (0.3cm and 1.0cm);
    \node[color=ucldpurple,anchor=east,at={(axis cs:56,700)}]{\textbf{This paper}}; 
    
    % puttnam2024_402
    \datapoint{puttnam2024_402}{(50,378.9)}{north east}{uclblue}{};
    \datapointgminocite{}{(50,402.2)}{}{uclblue};

    % puttnam2024_339
    \datapoint{puttnam2024_339}{(100,322.8)}{west}{uclblue}{};
    \datapointgminocite{}{(100,339.1)}{}{uclblue};
    
    % yang2026_135
    % O+SCL gmi/decoded: 201.39/190.33 + 135.51/127.28 = 336.9/317.61
    % Don't mark lower bound, only combined
    % \addplot[uclblue,only marks,mark=*,mark size=2pt,opacity=0.5,forget plot] coordinates {(39,317.61)};
    % \addplot[uclblue,only marks,mark=o,mark size=2pt,line width=1pt,opacity=0.5,forget plot] coordinates {(39,336.90)};
    % \node[color=uclblue,anchor=east,at={(axis cs:39,336.9)}]{\cite{yang2026_135}};
    
    % luis2025_430
    \datapoint{luis2025_430}{(10,398.6)}{south east}{uclblue}{\textsuperscript{$\ast$}};
    \datapointgminocite{}{(10,430.2)}{}{uclblue};

    % luis2026_450
    \datapoint{luis2026_450}{(39,418.75)}{south east}{uclblue}{};
    \datapointgminocite{}{(39,450.08)}{}{uclblue};
    
    % yang2025_300
    \datapoint{yang2025_300}{(39,275.82)}{north east}{uclblue}{};
    \datapointgminocite{}{(39,300.28)}{}{uclblue};
    
    % soma2024multiband
    % \datapoint{soma2024multiband}{(45,119.35)}{south west}{uclblue}{};

    % \datapointsq{ali2025unrepeated}{(4.4,25.6)}{south west}{uclred}{};
    
    % liu2026tb
    \draw[densely dotted](axis cs:100,188.8)--(axis cs:100,377.6);
    \datapointsqgmi{liu2026tb}{(100,188.8)}{south east}{uclred}{};
    \addplot[uclred,only marks,mark=square,mark size=2pt,line width=1pt,opacity=0.5,forget plot] coordinates {(100,377.6)};

    % % liu2024_502
    % \draw[densely dotted](axis cs:11.04,251.3)--(axis cs:11.04,502.6);
    % \datapointsqgmi{liu2024_502}{(11.04,251.3)}{west}{uclred}{};
    % \addplot[uclred,only marks,mark=square,mark size=2pt,line width=1pt,opacity=0.5,forget plot] coordinates {(11.04,502.6)};

    % li2026beyond
    \draw[densely dotted](axis cs:10.9,251.3)--(axis cs:10.9,550.97);
    \datapointsqgmi{li2026beyond}{(10.9,275.49)}{west}{uclred}{};
    \addplot[uclred,only marks,mark=square,mark size=2pt,line width=1pt,opacity=0.5,forget plot] coordinates {(10.9,550.97)};

    % zhang2025_273
    \draw[densely dotted](axis cs:1.4,136.8)--(axis cs:1.4,273.6);
    \datapointsq{zhang2025_273}{(1.4,136.8)}{west}{uclred}{};
    \addplot[uclred,only marks,mark=square*,mark size=2pt,line width=1pt,opacity=0.5,forget plot] coordinates {(1.4,273.6)};

    \addlegendentry{};
    \addlegendimage{uclblue,only marks,mark=o,mark size=2pt,line width=1pt};
    \addlegendentry{Silica fibre};
    \addlegendimage{uclblue,only marks,mark=*,mark size=2pt,line width=1pt};
    \addlegendentry{Aggregate Bi-Di throughput};
    \addlegendimage{densely dotted};
    \addlegendentry{};
    \addlegendimage{uclred,only marks,mark=square,mark size=2pt,line width=1pt};
    \addlegendentry{Hollow-core fibre};
    \addlegendimage{uclred,only marks,mark=square*,mark size=2pt,line width=1pt};
    \addlegendentry{Hybrid single-/multi-mode};
    \addlegendimage{uclblue,only marks,mark=asterisk};

    \end{groupplot}

    \end{tikzpicture}
    % \vspace{-2.5em}
    \caption{(a) Experimental UWB SMF transmission~\cite{puttnam2024_402,luis2026_450,luis2025_430,zhang2025_273,liu2026tb,li2026beyond,puttnam2024_339,yang2025_300}. Filled/open markers: net/GMI throughput.}
    \vspace{-.5em}
    \label{fig:progress}
\end{figure}

Here, we demonstrate Bi-Di OESCL-band HCF transmission for the first time, exploiting the low-nonlinearity, wide-bandwidth provided by the 1\textsuperscript{st} window fibre design, and low-backscattering properties of HCF. By combining HCF-specific digital signal processing (DSP) to combat gas-absorption alongside a record 42.5\,THz total bandwidth, enabled by novel bismuth-doped fibre amplifiers (BDFAs) developed to suppport $\sim$100\,nm O-band transmission, we achieved an aggregate Bi-Di GMI throughput of 423.7\,Tb/s\,+\,426.5\,Tb/s and a decoded rate of 396.9\,Tb/s\,+\,399.9\,Tb/s over a 60\,km HCF span. 
%By combining HCF-specific digital signal processing (DSP) to combat gas-absorption alongside novel bismuth-doped fibre amplifiers (BDFAs) to achieve $>$100\,nm O-band transmission and record 42.5\,THz total bandwidth, we achieve an aggregate Bi-Di GMI throughput of 850.24\,Tb/s (423.73\,Tb/s\,+\,426.51\,Tb/s) and decoded rate of 796.77\,Tb/s (396.92\,Tb/s\,+\,399.85\,Tb/s) over 60\,km HCF span. 
As shown in Fig.~\ref{fig:progress}(a), these results represent %the highest data-rate reported in any purely single-mode fibre in both directions simultaneously, 
the widest transmission bandwidth %in any optical fibre transmission system, 
and the highest aggregate capacity-distance product (51\,Pb/s$\cdot$km Bi-Di) reported in a single-span single-mode fibre transmission system, demonstrating the potential of HCF systems to underpin future high-capacity, UWB optical networks.

\vspace{-.5em}
\section{Experimental setup}

\begin{figure*}[t]
    \centering
    \includegraphics[width=\linewidth]{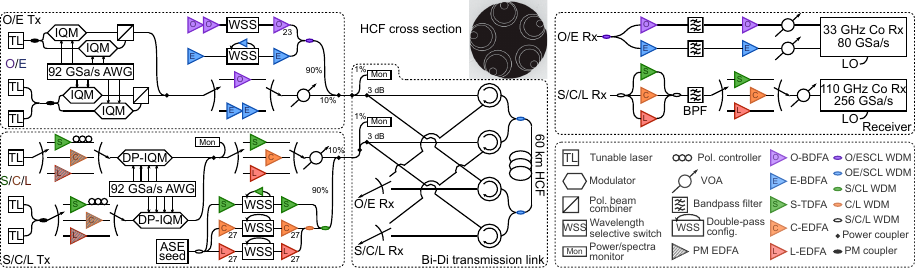}
    \caption{Experimental diagram of Bi-Di HCF transmission. Numbers below the amplifiers: maximum powers (in dBm) of the high-power amplifiers. Inset: cross section of the HCF.}
    \label{fig:diagram}
    \vspace{-.8em}
\end{figure*}

Figure~\ref{fig:diagram} shows the UWB Bi-Di transmission setup. In the O-/E-band transmitter (Tx), a 200\,kHz-linewidth tunable laser (TL) and two independent single-polarisation in-phase/quadrature modulators (IQMs), each driven by a 92\,GSa/s arbitrary waveform generator (AWG), were used to generate dual-polarisation (DP) root-raised-cosine-shaped signals with 1\,\% roll-off via a polarisation beam combiner (PBC). In the S/C/L-bands, dual-polarisation IQMs (DP-IQMs) were employed with $<$10\,kHz laser linewidth. The centre channel under test (CUT) was combined with two neighbouring channels and amplified to form a 3×32\,GBaud sliding test band with 1.33\,GHz guard-band, based on a 33.33\,GHz frequency grid. Dummy WDM signals were generated from spectrally-shaped amplified spontaneous emission (ASE) using commercial O/S/C/L-band wavelength selective switches (WSSs) or a custom-developed E-band WSS~\cite{fontaine2023multiport}, operated in double-pass configurations. The amplifiers used in this experiment were in-house built BDFAs for the O/E-bands~\cite{mikhailov2024o,donodin2025high} (typical output power of 20\,dBm), thulium-doped fibre amplifiers (TDFA) for the S-band (20\,dBm), and erbium-doped fibre amplifiers (EDFA) for the C/L-band (23\,dBm). The values indicated below the amplifiers in Fig.~\ref{fig:diagram} denote the maximum output powers of the high-power amplifier units used to generate dummy channels. 

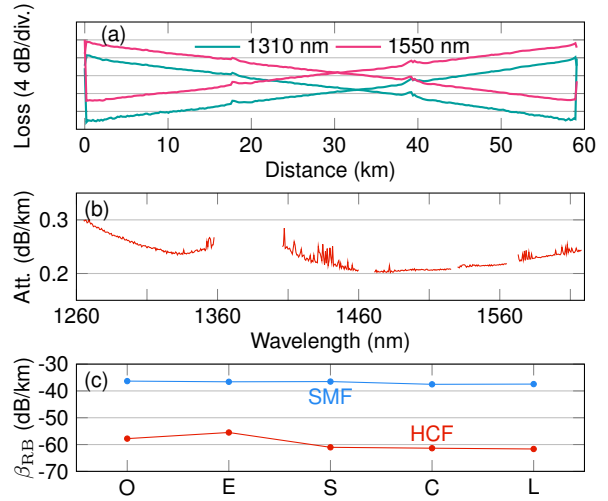
\begin{figure}[t]
    \centering
    % \vspace{-1em}
    \begin{tikzpicture}\footnotesize
      \hypersetup{hidelinks}
      \begin{groupplot}[
          	group style={
    		group size=1 by 3,
    		x descriptions at=all,
    		y descriptions at=all,
    		vertical sep=3em,
    	},
    height=3cm,
    width=8.3cm,
    ymajorgrids=true,
    ylabel style={at={(axis description cs:.08,.5)}},
    clip=false,
    ]

    \nextgroupplot[
        xlabel={Distance (km)},
        ylabel={Loss (4 dB/div.)},
        xmin=-1,xmax=60,
        ymin=4,ymax=28,
        yticklabels=\empty,
        xtick={0,10,...,60},
        xticklabels={0,10,...,60},
        ytick={24,20,...,8},
        legend style={anchor=north,at={(axis cs: 30,27)},draw=none,fill=none,row sep=-2pt},
        legend columns=2,
        xlabel style={at={(axis description cs:.5,.14)}},
        clip=false,
    ]

    \addlegendentry{1310~nm};
    \addlegendimage{uclg9,thick};
    
    \addlegendentry{1550~nm};
    \addlegendimage{uclg8,thick};
    
    \addplot[uclg9,thick,each nth point=5] table[x=port1_1310_distance,y=port1_1310_trace]{data/OTDR_results_truncated2.txt};
    \addplot[uclg9,thick,each nth point=5] table[x=port2_1310_distance,y=port2_1310_trace]{data/OTDR_results_truncated2.txt};
    \addplot[uclg8,thick,each nth point=5] table[x=port1_1550_distance,y=port1_1550_trace]{data/OTDR_results_truncated2.txt};
    \addplot[uclg8,thick,each nth point=5] table[x=port2_1550_distance,y=port2_1550_trace]{data/OTDR_results_truncated2.txt};

    \node[at={(axis cs:1,29)},anchor=north west]{(a)};

    \nextgroupplot[
        xlabel={Wavelength (nm)},
        ylabel={Att. (dB/km)},
        xmin=1260,xmax=1620,
        ymin=0.15,ymax=0.35,
        xtick={1260,1310,...,1610},
        xticklabels={1260,,1360,,1460,,1560,},
        ytick={0.2,0.3},
        yticklabels={0.2,0.3},
        xlabel style={at={(axis description cs:.5,.14)}},
        clip=false,
        unbounded coords=jump,
    ]
    \addplot[uclred] table[x=wavelength,y=loss]{data/paper_loss_hcf.txt};

    \node[at={(axis cs:1260,0.35)},anchor=north west]{(b)};

    \nextgroupplot[
    xmin=0.5,xmax=5.5,
    ymin=-70, ymax=-30,
    xtick={1,2,3,4,5},
    xticklabels={{O},{E},{S},{C},{L}},
    ytick={-70,-60,-50,-40,-30},
    yticklabels={-70,-60,-50,-40,-30},
    xlabel=\empty,
    ylabel={$\beta_\mathrm{RB}$ (dB/km)},
    % title={$\beta_\mathrm{RB}$ (dB/km)},
    % title style={at={(axis cs: 3,-24)},anchor=north},
    % axis background/.style={fill=white},
    ]
    \addplot[uclblue,mark=*,mark size=1pt] table[x=band,y=bs_smf]{data/paper_backscattering.txt};
    \addplot[uclred,mark=*,mark size=1pt] table[x=band,y=bs_hcf]{data/paper_backscattering.txt};
    \node[at={(axis cs:3,-42)}]{\textcolor{uclblue}{SMF}};
    \node[at={(axis cs:4,-56)}]{\textcolor{uclred}{HCF}};

    \node[at={(axis cs:0.5,-30)},anchor=north west]{(c)};

    \end{groupplot}

    \end{tikzpicture}
    \vspace{-1.5em}
    \caption{Fibre characterisations. (a) OTDR traces. (b) Fibre attenuation. (c) Measured RB coefficient.}
    \label{fig:fibre}
    \vspace{-.5em}
\end{figure}

The Bi-Di transmission link comprised two 3-dB couplers to split the WDM signals into two copies for forward (FW) and backward (BW) transmission, four optical circulators to direct the Tx and receiver (Rx) signal, and a 60\,km HCF designed to operate in the fundamental transmission window, similar to those reported in~\cite{petrovich2025broadband,ali2025unrepeated,hong2025real}. The cross section of the HCF is shown in the inset of Fig.~\ref{fig:diagram}. Figure~\ref{fig:fibre} shows the measured fibre parameters, including optical time-domain reflectometry (OTDR) traces measured from both ends of this fibre span, wavelength-dependent HCF attenuation, and Rayleigh backscattering coefficient, which is more than 20\,dB lower than that of silica SMF. The inter-modal interference was measured to be $<$-56\,dB/km between 1450\,nm and 1620\,nm. To achieve high port 1$\to$3 directivity ($>$50\,dB), separate circulators were used for O/E-bands and S/C/L-bands which were operated on parallel branches and combined or split at each fibre end. 

The receiver path of each direction was amplified and measured in each band separately. The CUT was filtered with an optical bandpass filter (BPF), attenuated with a variable optical attenuator (VOA), and detected by a coherent (Co) Rx using a local oscillator (LO) of the same specification as the transmitter lasers. 80\,GSa/s and 256\,GSa/s real-time oscilloscopes were used to acquire the signals in the O/E- and S/C/L-bands, respectively. Pilot-based DSP with 4\,\% overhead, adaptive rate decoding, and code rate puncturing (with granularity of $\sim$\,0.01) were applied offline~\cite{yang2025transmission}. 

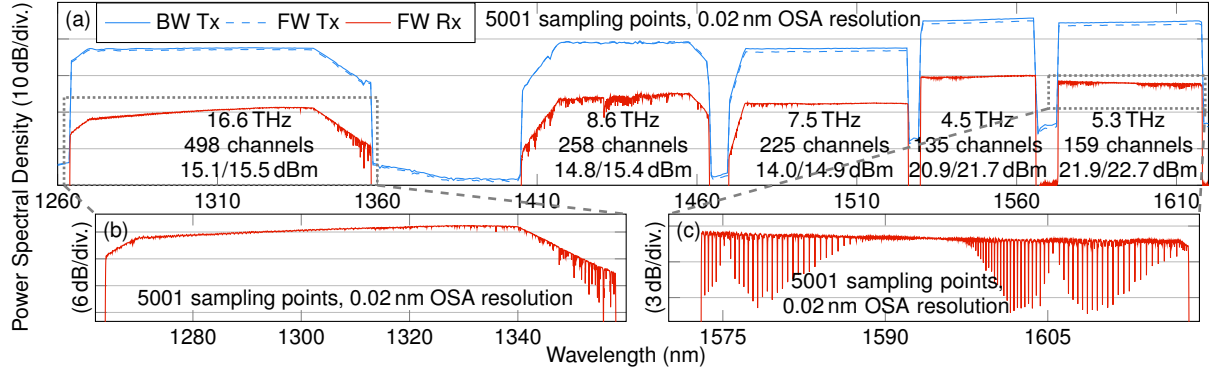
\begin{figure*}[ht]

% Left plot zoom
\newcommand\zoomxa{1400}
\newcommand\zoomxb{1467}
\newcommand\zoomya{-45}
\newcommand\zoomyb{-72}

% Right plot zoom
\newcommand\zoomxc{1570}
\newcommand\zoomxd{1619}
\newcommand\zoomyc{-42}
\newcommand\zoomyd{-51}

% O-band zoom
\newcommand\zoomxe{1262}
\newcommand\zoomxf{1360}
\newcommand\zoomye{-48}
\newcommand\zoomyf{-72}
% \vspace{-1em}
    \centering
    \pgfplotsset{very non boxed y axis/.style={}}
    \begin{tikzpicture}\footnotesize
    \begin{groupplot}[group style={group size=1 by 3,xticklabels at=edge bottom,horizontal sep=0pt,vertical sep=0pt},
    height=4.0cm,
    ]
    
    \nextgroupplot[
    width=1.05\linewidth,
    xmin=1260, xmax=1620,
    ymax=-22,ymin=-72,
    xtick={1260,1310,...,1610},
    xticklabels={1260,1310,...,1610},
    ytick={-72,-62,-52,-42,-32},
    yticklabels=\empty,
    ylabel={Power Spectral Density (10\,dB/div.)},
    ylabel style={at={(axis description cs: .05,.1)}},
    ymajorgrids=true,
    legend style ={anchor=north west,at={(axis cs: 1272, -22)},row sep=-3pt, legend columns=3},  % fill=none, draw=none,
    legend cell align={left}, 
    ]
    \addplot[uclblue,forget plot,each nth point=10] table[x=wavelength,y=power_bwtx]{data/paper_spec.txt};
    \addplot[uclblue,dashed,forget plot,each nth point=10] table[x=wavelength,y=power_fwtx]{data/paper_spec.txt};
    \addplot[uclred,forget plot] table[x=wavelength,y=power_fwrx]{data/paper_spec.txt};

    \node[align=center,text width=3cm,anchor=south] at (axis cs:1320,-72){16.6\,THz\\498 channels\\15.1/15.5\,dBm};

    \coordinate (c0) at (axis cs:\zoomxe,\zoomye);
    \coordinate (c1) at (axis cs:\zoomxe,\zoomyf);
    \coordinate (c2) at (axis cs:\zoomxf,\zoomye);
    \coordinate (c3) at (axis cs:\zoomxf,\zoomyf);
    \draw [gray,densely dotted,line width=1pt] (c0) rectangle (c3);

    \coordinate (b0) at (axis cs:\zoomxc,\zoomyc);
    \coordinate (b1) at (axis cs:\zoomxc,\zoomyd);
    \coordinate (b2) at (axis cs:\zoomxd,\zoomyc);
    \coordinate (b3) at (axis cs:\zoomxd,\zoomyd);

    \draw [gray,densely dotted,line width=1pt] (b0) rectangle (b3);

    \addlegendentry{BW Tx};
    \addlegendimage{uclblue};
    \addlegendentry{FW Tx};
    \addlegendimage{uclblue,dashed};
    \addlegendentry{FW Rx};
    \addlegendimage{uclred};

    \node[align=center,text width=3cm,anchor=south] at (axis cs:1437,-72){8.6\,THz\\258 channels\\14.8/15.4\,dBm};
    \node[align=center,text width=3cm,anchor=south] at (axis cs:1500,-72){7.5\,THz\\225 channels\\14.0/14.9\,dBm};
    \node[align=center,text width=3cm,anchor=south] at (axis cs:1548,-72){4.5\,THz\\135 channels\\20.9/21.7\,dBm};
    \node[align=center,text width=3cm,anchor=south] at (axis cs:1595,-72){5.3\,THz\\159 channels\\21.9/22.7\,dBm};

    \node at (axis cs:1462,-27){5001 sampling points, 0.02\,nm OSA resolution};

    \nextgroupplot[
        anchor=south,
        at={(group c1r1.south)},
        height=3cm,
        width=8.6cm,
        yshift=-1.8cm,
        xshift=-3.6cm,
        xmin=\zoomxe, xmax=\zoomxf,
        ymax=\zoomye,ymin=\zoomyf,
        ymajorgrids=true,
        xtick={1260,1280,...,1340},
        xticklabels={1260,1280,...,1340},
        ytick={-52,-58,...,-70},
        yticklabels=\empty,
        xlabel={Wavelength (nm)},
        xlabel style={at={(axis cs: 1360,-67)}},
        ylabel={(6\,dB/div.)},
        ylabel style={at={(axis description cs: .15,0.5)}},
    ]
    \addplot[uclred,forget plot] table[x=wavelength_o,y=power_o]{data/paper_spec.txt};
    \node at (axis cs:1310,-67){5001 sampling points, 0.02\,nm OSA resolution};

    \nextgroupplot[
    anchor=west,
    at={(group c1r2.east)},
    height=3cm,
    width=8.6cm,
    xshift=2em,
    xmin=\zoomxc, xmax=\zoomxd,
    ymax=\zoomyc,ymin=\zoomyd,
    xtick={1575,1590,1605,1620},
    xticklabels={1575,1590,1605,1620},
    ytick={-49,-46,-43},
    yticklabels=\empty,
    ylabel={(3\,dB/div.)},
    ylabel style={at={(axis description cs: .15,0.5)}},
    ymajorgrids=true,
    ]
    
    \addplot[uclred,forget plot] table[x=wavelength_l,y=power_l]{data/paper_spec.txt};
    \node[align=center,text width=3cm,anchor=south] at (axis cs:1591,-51){5001 sampling points,\\0.02\,nm OSA resolution};
    \end{groupplot}

    \draw [gray,dashed,line width=1pt] (b1) -- (group c1r3.north west);
    \draw [gray,dashed,line width=1pt] (b3) -- (group c1r3.north east);
    \draw [gray,dashed,line width=1pt] (c1) -- (group c1r2.north west);
    \draw [gray,dashed,line width=1pt] (c3) -- (group c1r2.north east);

    \node[at={(group c1r1.north west)+(1cm,1cm)},anchor=north west]{(a)};
    \node[at={(group c1r2.north west)+(1cm,1cm)},anchor=north west]{(b)};
    \node[at={(group c1r3.north west)+(1cm,1cm)},anchor=north west]{(c)};
    
    \end{tikzpicture}
    % \vspace{-1em}
    \caption{(a) Transmitted FW/BW and received FW spectra, and signal characteristics of bandwidth, number of channels, and FW/BW launch power. (b) O-band received (zoom). (c) L-band received (zoom).}
    \label{fig:spec}
    \vspace{-.5em}
\end{figure*}

\section{Transmission results}
The signal spectra before and after transmission, measured using an optical spectrum analyser (OSA), are shown in Fig.~\ref{fig:spec}(a). The WDM signal characterisations are also summarised in Fig.~\ref{fig:spec}(a). A launch power (LP) difference between the two directions of less than 1\,dB was achieved across all bands. A wavelength-dependent link loss of 14-19\,dB across the five bands was measured, with gas line absorption (GLA) or water absorption present at certain wavelengths. The O-band exhibited the widest usable bandwidth and a flat attenuation profile over more than 80\,nm, as shown in Fig.~\ref{fig:spec}(a), while in contast, the L-band was visibly affected by CO\textsubscript{2} absorption as seen in the zoomed spectrum in Fig.~\ref{fig:spec}(c). Both gas and water absorption peaks were observed across the E-band spectrum. For these channels, adaptive MIMO filter lengths of up to 163 taps, carrier phase recovery averaging up to 46 pilot symbols, and an additional GLA compensation step~\cite{sillekens2026gas} were applied in offline DSP to improve performance.

\begin{figure*}[b]
    \vspace{-2em}
    \centering
    \pgfplotsset{very non boxed y axis/.style={}}
    \begin{tikzpicture}\footnotesize
    \begin{groupplot}[group style={group size=1 by 2,xticklabels at=edge bottom,horizontal sep=0pt,vertical sep=0pt},
    height=3.2cm,width=1.04\linewidth,
    ]

    \nextgroupplot[
    xmin=1260, xmax=1620,
    ymax=25,ymin=0,
    ymajorgrids=true,
    xtick={1260,1310,...,1610},
    xticklabels=\empty,
    ytick={5,10,15,20,25},
    ylabel={SNR (dB)},
    ylabel style={at={(axis description cs: .035,.5)}},
    yticklabels={5,10,15,20,25},
    only marks,
    legend style ={anchor=south,at={(axis cs: 1380, 5)},row sep=-3pt},
    legend cell align={left},
    ]
    \addplot[mark=o,mark size=1pt,opacity=0.5,forget plot,restrict y to domain=2.5:inf] table[x=wavelength,y=snr_fw]{data/paper_result_o.txt};
    \addplot[uclred,mark=triangle,mark size=.8pt,opacity=0.5,forget plot,restrict y to domain=2.5:inf] table[x=wavelength,y=snr_bw]{data/paper_result_o.txt};
    \addplot[uclgreen,mark=square,mark size=1pt,line width=.8pt,forget plot] table[x=wavelength,y=snr_unidi]{data/paper_result_unidi.txt};
    
    \addplot[mark=o,mark size=1pt,opacity=0.5,forget plot,restrict y to domain=2.5:inf] table[x=wavelength,y=snr_fw]{data/paper_result_e.txt};
    \addplot[mark=o,mark size=1pt,opacity=0.5,forget plot] table[x=wavelength,y=snr_fw]{data/paper_result_s.txt};
    \addplot[mark=o,mark size=1pt,opacity=0.5,forget plot] table[x=wavelength,y=snr_fw]{data/paper_result_c.txt};
    \addplot[mark=o,mark size=1pt,opacity=0.5,forget plot] table[x=wavelength,y=snr_fw]{data/paper_result_l.txt};
    
    \addplot[uclred,mark=triangle,mark size=.8pt,opacity=0.5,forget plot,restrict y to domain=2.5:inf] table[x=wavelength,y=snr_bw]{data/paper_result_e.txt};
    \addplot[uclred,mark=triangle,mark size=.8pt,opacity=0.5,forget plot] table[x=wavelength,y=snr_bw]{data/paper_result_s.txt};
    \addplot[uclred,mark=triangle,mark size=.8pt,opacity=0.5,forget plot] table[x=wavelength,y=snr_bw]{data/paper_result_c.txt};
    \addplot[uclred,mark=triangle,mark size=.8pt,opacity=0.5,forget plot] table[x=wavelength,y=snr_bw]{data/paper_result_l.txt};

    \addplot[uclgreen,mark=square,mark size=1pt,line width=.8pt,forget plot, restrict x to domain=1390:1620] table[x=wavelength,y=snr_unidi]{data/paper_result_unidi.txt};

    \addlegendentry{FW};
    \addlegendimage{only marks,mark=o,mark size=1.2pt};
    \addlegendentry{BW};
    \addlegendimage{only marks,uclred,mark=triangle,mark size=1.2pt};
    \addlegendentry{Uni-Di};
    \addlegendimage{uclgreen,mark=square,mark size=1pt,line width=1pt};

    \nextgroupplot[
    height=4.0cm,
    xmin=1260, xmax=1620,
    ymin=00, ymax=500,
    xtick={1260,1310,...,1610},
    xticklabels={1260,1310,...,1610},
    ytick={0,100,200,300,400},
    yticklabels={0,0.1,0.2,0.3,0.4},
    ylabel=\empty,
    xlabel={Wavelength (nm)},
    xlabel style={at={(axis cs: 1430,40)}},
    ylabel={Data rate (Tb/s)},
    ylabel style={at={(axis description cs: .035,.5)}},
    ymajorgrids=true,
    clip=false,
    only marks,
    scatter, 
    scatter src=explicit,
    colormap={my4}{
    rgb255(0cm)=(154, 59, 255)
    rgb255(1cm)=(36, 136, 242)
    rgb255(2cm)=(47, 148, 26)
    rgb255(3cm)=(255, 118, 38)
    rgb255(4cm)=(0, 0, 0)
    },
    point meta min=1, point meta max=5,
    scatter/use mapped color={
        draw=mapped color,
        fill=mapped color
    },
    legend style ={anchor=south,at={(axis cs: 1540, -10)},fill=none,draw=none,row sep=-3pt},
    legend columns = 2,
    legend cell align={left},
    ]
    \addplot[mark=o,mark size=1pt,opacity=0.3,forget plot,point meta=5,restrict y to domain=50:inf] table[x=wavelength,y=datarate_fw]{data/paper_result_o.txt};
    \addplot[mark=triangle,mark size=.8pt,opacity=0.3,forget plot,point meta=5,restrict y to domain=50:inf] table[x=wavelength,y=datarate_bw]{data/paper_result_o.txt};
    
    \addplot[mark=*,mark size=.5pt,forget plot,restrict y to domain=50:inf] table[x=wavelength,y=netrate_fw,meta=modfmt_fw]{data/paper_result_o.txt};
    \addplot[mark=triangle*,mark size=.5pt,forget plot,restrict y to domain=50:inf] table[x=wavelength,y=netrate_bw,meta=modfmt_bw]{data/paper_result_o.txt};

    \node[at={(axis cs:1290,350)},anchor=south]{\includegraphics[width=1.1cm]{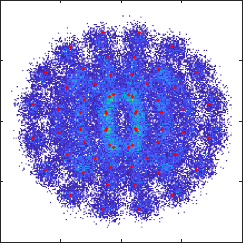}};
    \draw[>=stealth,->](axis cs:1329.7,300)--(axis cs:1303,390);
    \node[at={(axis cs:1330,400)},anchor=south]{\includegraphics[width=1.1cm]{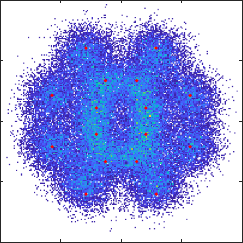}};
    \draw[>=stealth,->](axis cs:1351.2,200)--(axis cs:1342,420);

    \node[align=center,text width=3cm,anchor=south] at (axis cs:1310,0){\textcolor{uclred}{269.0 (GMI)\\252.8 (FEC)}};

    \addplot[mark=o,mark size=1pt,opacity=0.3,forget plot,point meta=5,restrict y to domain=50:inf] table[x=wavelength,y=datarate_fw]{data/paper_result_e.txt};
    \addplot[mark=triangle,mark size=.8pt,opacity=0.3,forget plot,point meta=5,restrict y to domain=50:inf] table[x=wavelength,y=datarate_bw]{data/paper_result_e.txt};
    \addplot[mark=o,mark size=1pt,opacity=0.3,forget plot,point meta=5] table[x=wavelength,y=datarate_fw]{data/paper_result_s.txt};
    \addplot[mark=triangle,mark size=.8pt,opacity=0.3,forget plot,point meta=5] table[x=wavelength,y=datarate_bw]{data/paper_result_s.txt};
    \addplot[mark=o,mark size=1pt,opacity=0.3,forget plot,point meta=5] table[x=wavelength,y=datarate_fw]{data/paper_result_c.txt};
    \addplot[mark=triangle,mark size=.8pt,opacity=0.3,forget plot,point meta=5] table[x=wavelength,y=datarate_bw]{data/paper_result_c.txt};
    \addplot[mark=o,mark size=1pt,opacity=0.3,forget plot,point meta=5] table[x=wavelength,y=datarate_fw]{data/paper_result_l.txt};
    \addplot[mark=triangle,mark size=.8pt,opacity=0.3,forget plot,point meta=5] table[x=wavelength,y=datarate_bw]{data/paper_result_l.txt};

    \addplot[mark=*,mark size=.5pt,forget plot,restrict y to domain=50:inf] table[x=wavelength,y=netrate_fw,meta=modfmt_fw]{data/paper_result_e.txt};
    \addplot[mark=triangle*,mark size=.5pt,forget plot,restrict y to domain=50:inf] table[x=wavelength,y=netrate_bw,meta=modfmt_bw]{data/paper_result_e.txt};
    \addplot[mark=*,mark size=.5pt,forget plot] table[x=wavelength,y=netrate_fw,meta=modfmt_fw]{data/paper_result_s.txt};
    \addplot[mark=triangle*,mark size=.5pt,forget plot] table[x=wavelength,y=netrate_bw,meta=modfmt_bw]{data/paper_result_s.txt};
    \addplot[mark=*,mark size=.5pt,forget plot] table[x=wavelength,y=netrate_fw,meta=modfmt_fw]{data/paper_result_c.txt};
    \addplot[mark=triangle*,mark size=.5pt,forget plot] table[x=wavelength,y=netrate_bw,meta=modfmt_bw]{data/paper_result_c.txt};
    \addplot[mark=*,mark size=.5pt,forget plot] table[x=wavelength,y=netrate_fw,meta=modfmt_fw]{data/paper_result_l.txt};
    \addplot[mark=triangle*,mark size=.5pt,forget plot] table[x=wavelength,y=netrate_bw,meta=modfmt_bw]{data/paper_result_l.txt};

    \addlegendentry{};
    \addlegendimage{black,only marks,mark=o,mark size=1.2pt};
    \addlegendentry{GMI};
    \addlegendimage{black,only marks,mark=triangle,mark size=1.2pt};
    \addlegendentry{};
    \addlegendimage{uclpurple,only marks,mark=*,mark size=1.2pt};
    \addlegendentry{Dec. GS-16\,QAM};
    \addlegendimage{uclpurple,only marks,mark=triangle*,mark size=1.2pt};
    \addlegendentry{};
    \addlegendimage{uclblue,only marks,mark=*,mark size=1.2pt};
    \addlegendentry{Dec. GS-64\,QAM};
    \addlegendimage{uclblue,only marks,mark=triangle*,mark size=1.2pt};
    \addlegendentry{};
    \addlegendimage{uclgreen,only marks,mark=*,mark size=1.2pt};
    \addlegendentry{Dec. GS-256\,QAM};
    \addlegendimage{uclgreen,only marks,mark=triangle*,mark size=1.2pt};
    \addlegendentry{};
    \addlegendimage{uclorange,only marks,mark=*,mark size=1.2pt};
    \addlegendentry{Dec. GS-1024\,QAM};
    \addlegendimage{uclorange,only marks,mark=triangle*,mark size=1.2pt};

    \node[at={(axis cs:1490,450)},anchor=south]{\includegraphics[width=1.1cm]{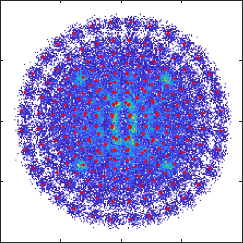}};
    \draw[>=stealth,->](axis cs:1513.3,400)--(axis cs:1503,480);
    \node[at={(axis cs:1550,500)},anchor=south]{\includegraphics[width=1.1cm]{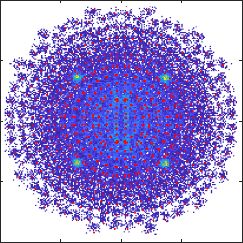}};
    \draw[>=stealth,->](axis cs:1548.0,470)--(axis cs:1550,520);

    \node[align=center,text width=3cm,anchor=south] at (axis cs:1430,0){\textcolor{uclred}{161.9 (GMI)\\149.1 (FEC)}};
    \node[align=center,text width=3cm,anchor=south] at (axis cs:1482,0){\textcolor{uclred}{169.4 (GMI)\\160.1 (FEC)}};
    \node[align=center,text width=3cm,anchor=south] at (axis cs:1548,230){\textcolor{uclred}{123.9 (GMI)\\116.8 (FEC)}};
    \node[align=center,text width=3cm,anchor=south] at (axis cs:1598,0){\textcolor{uclred}{126.0 (GMI)\\118.0 (FEC)}};

    % \nextgroupplot[
    % at={(group c1r2.west)},
    % xshift=5.3cm,
    % yshift=0.8cm,
    % height=3cm,
    % width=3cm,
    % xmin=0.5,xmax=5.5,
    % ymin=-70, ymax=-30,
    % xtick={1,2,3,4,5},
    % xticklabels={{O},{E},{S},{C},{L}},
    % ytick={-70,-60,-50,-40,-30},
    % yticklabels={-70,-60,-50,-40,-30},
    % xlabel=\empty,
    % ylabel=\empty,
    % title={$\beta_\mathrm{RB}$ (dB/km)},
    % title style={at={(axis cs: 3,-24)},anchor=north},
    % axis background/.style={fill=white},
    % ]
    % \addplot[uclblue,mark=*,mark size=1pt] table[x=band,y=bs_smf]{data/paper_backscattering.txt};
    % \addplot[uclred,mark=*,mark size=1pt] table[x=band,y=bs_hcf]{data/paper_backscattering.txt};
    % \node[at={(axis cs:3,-42)}]{\textcolor{uclblue}{SMF}};
    % \node[at={(axis cs:4,-56)}]{\textcolor{uclred}{HCF}};
    
    \end{groupplot}
    % \node[at={(group c1r1.north west)},xshift=.4cm,yshift=-.3cm]{\large{(a)}};
    % \node[at={(group c1r2.north west)},xshift=.4cm,yshift=-.4cm]{\large{(b)}};

    \node[at={(group c1r1.north west)+(1cm,1cm)},anchor=north west]{(a)};
    \node[at={(group c1r2.north west)+(1cm,1cm)},anchor=north west]{(b)};

    \end{tikzpicture}
    % \vspace{-2.5em}
    \caption{Measured (a) SNR (b) GMI and decoded data rate after deducting pilot overhead. Inset: constellation diagrams and per-band aggregate bi-directional throughputs (Tb/s). }
    \label{fig:result}
    % \vspace{-1.8em}
\end{figure*}
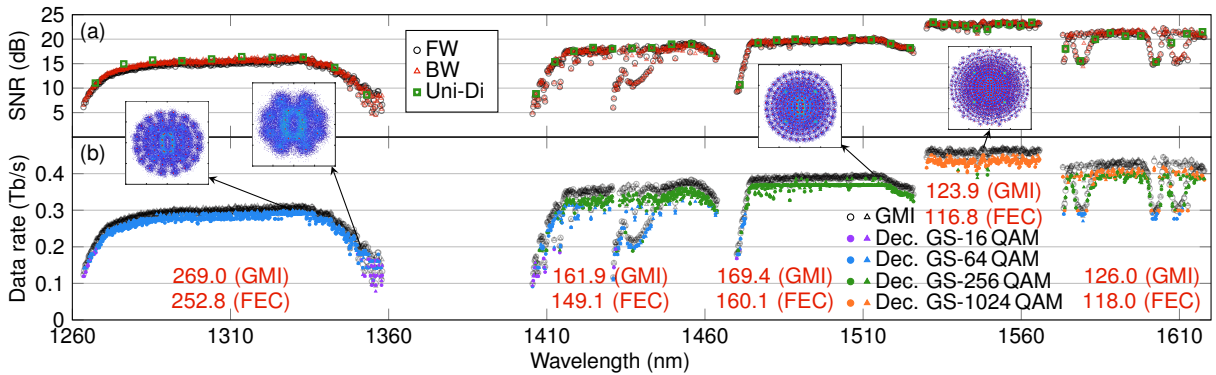

Figure~\ref{fig:result}(a) shows the SNR of the measured 1275 channels spanning 42.5\,THz optical bandwidth for both transmission directions, obtained by averaging the best three of five traces obtained per wavelength channel. Nine edge channels in the O- and E-bands were excluded due to poor transceiver performance or water absorption distortion. The reduced SNR observed in the centres of E- and L-bands corresponds to the water/gas absorption lines shown in Fig.~\ref{fig:spec}(c). In contrast, the O-band exhibits uniform SNR across the flat LP region and a gradual performance decrease toward both edges as the LP rolls off, enabling near-complete use of its 100\,nm spectral range. Figure~\ref{fig:spec}(a)(b) shows how the custom BDFA design enables almost as many channels as the SCL-bands combined. SNRs for uni-directional (Uni-Di) transmission over the same fibre link with identical launch powers were measured for 10 channels in each band and are plotted as green square markers in Fig.\ref{fig:result}(a). %The measured Rayleigh backscattering coefficient for each band, plotted in the inset of Fig.~\ref{fig:result}(b), is more than 20\,dB lower than that of silica SMF. 
The average SNR penalty due to Rayleigh backscattering in same-wavelength Bi-Di transmission is less than 0.32\,dB, demonstrating the suitability of HCF for UWB Bi-Di transmission. %This is attributed to the approximately \X{20???}\,dB/km reduction in backscattering~\cite{} compared with conventional silica fibres. 
%This is attributed to the reduced backscattering coefficient compared with conventional silica fibres.

Different geometrically-shaped constellations with four cardinalities (GS-16\,QAM, GS-64\,QAM, GS-256\,QAM, GS-1024\,QAM)~\cite{sillekens2022high}, plotted in the inset of Fig.~\ref{fig:result}, were tested for each channel, and the one yielding the highest decoded data rate was selected and plotted with coloured markers in Fig.~\ref{fig:result}(b), while black markers indicate the data rate estimated from GMI. Despite a higher fibre attenuation ($\sim$0.06\,dB/km) and $\sim$4\,dB lower back-to-back transceiver SNR, the O-band contributes the largest share of the total throughput owing to its widest usable bandwidth and the near-absence of absorption lines. The combined OESCL-band decoded data rate was 396.9\,Tb/s\,+\,399.9\,Tb/s for the two directions, with the GMI estimated data rate reaching 423.7\,Tb/s + 426.5\,Tb/s. These results mark the highest throughput achieved in any single-mode fibre in bi-directional configurations, showing that HCF is suitable for UWB Bi-Di transmission systems.
We note that our HCF attenuation is $>$0.2\,dB/km in this work, current state-of-the-art levels ($\sim$0.09\,dB/km) could make 200\,km ultra-high capacity single-span Bi-Di links feasible~\cite{poggiolini2025case}.

\section{Conclusion}
We have achieved a record aggregate bi-directional throughput of 423.7\,+\,426.5\,Tb/s (396.9\,+\,399.9\,Tb/s after decoding) in the same-wavelength bi-directional transmission, over 60\,km HCF using a 42.5\,THz signal bandwidth, custom HCF DSP and wideband BDFAs. These results show that in addition to low-loss, low-nonlinearity, and negligible SRS, HCFs, particularly those using the fundamental transmission window, can offer ultra-wide bandwidth and low backscattering levels suitable for high data-rate Bi-Di operation, potentially doubling the overall capacity of HCF cables in future ultra-high capacity, low-latency optical networks.

%-------------------------------------------------- Acknowledgements Section -------------------------------------------------------%
\clearpage
\section{Acknowledgements}
This work was funded by EPSRC grants EP/R035342/1 TRANSNET, EP/W015714/1 EWOC, EP/V007734/1. The authors would like to acknowledge Microsoft Azure Fiber (for HCF) and Lightera Fiber (for BDF) manufacturing teams for their support. RA is supported by a UCL Research Excellence Studentship, ES and AD by the Department for Science, Innovation and Technology (DSIT) and the Royal Academy of Engineering (RAEng) under Research Fellowships, and PB by a Royal Society Research Professorship.

%-------------------------------------------------- Bibliography Section -------------------------------------------------------%
% see also https://tex.stackexchange.com/questions/55030/text-before-references-but-after-bibliography-title-with-bibtex as of 2024-02-29
\defbibnote{myprenote}{%
% Citations must be easy and quick to find. More precisely:
% \begin{itemize}
%     \item Please list all the authors. 
%     \item The title must be given in full length. 
%     \item Journal and conference names should not be abbreviated but rather given in full length.
%     \item The DOI number should be added incl. a link.
% \end{itemize}
}
\printbibliography[prenote=myprenote]
% \bibliographystyle{IEEEtran}
% \bibliography{references}
\vspace{-4mm}

%%%%%%%%%%%%%%%%%%%%%%%%%%%%%%%%%%%%%%%%%%%%%
%---------------------------------------------- End of Document -----------------------------------------------%
\end{document}